\documentclass[aip,rsi,reprint,graphicx]{revtex4-1} 
\usepackage{graphicx}
\usepackage{bm, amssymb}
\usepackage{dcolumn}
\begin{document}

\title{A setup for Seebeck coefficient measurement through controlled heat pulses}
\author{Abdul Ahad}
\address{Department of Physics, Aligarh Muslim University, Aligarh 202002, India}
\author{D. K. Shukla}
\thanks{Corresponding Author: dkshukla@csr.res.in}
\address{UGC-DAE Consortium for Scientific Research, Indore 452001, India}

\date{\today}

\begin{abstract}
A setup is designed for measuring the Seebeck coefficient (S) of materials in form of thin film, bar and wire. The main feature of this setup is control in heating and cooling cycles. In this setup heat pulse is used to generate the temperature gradient. To demonstrate the capabilities of this setup, S vs T of standard wire samples such as Au-Fe (0.07 \%), chromel, Pt and thin films of Pt and F doped SnO$_2$ are presented. The standard uncertainty of the repeatability in S measurement is found to be $\sim$ $\pm 0.056~\mu V/K$ while temperature stability is $\sim$ $\pm 10~mK$ (at 320 K), estimated for a chromel wire sample. We have tested the setup in temperature range 100 K to 320 K, while it does not have any intrinsic limitation in going down to liquid He temperatures. For temperatures above 320 K limitation is due to gluing material like varnish. 
\end{abstract}

\maketitle

The urgency for thermoelectric materials that can convert heat into electricity is high, because of global warming and limited natural energy sources. For an effective thermoelectric material the correlation between the S, electrical conductivity ($\sigma$) and thermal conductivity ($\kappa$) plays an important role~\cite{Dre2007}. These all together decides the efficiency of the material for practical use through the relation~\cite{Sny2008} ZT = S$^2\sigma$T/$\kappa$, where ZT is figure of merit. Apart from this, the measurement of the S is useful to study the nature of mobile charge carriers (electron or holes), by the sign of S. It is good tool also for the studies of phase transitions, idea of density of states and electron-magnon or electron-phonon interactions~\cite{Abd2017,Zim1979} etc. Development of the S measurement system has always been resurgent among the researchers. Some reports focus on the sample mounting procedure, like sandwich mounting~\cite{Mul2016,Bof2005}, etc. while some focus on the bridging mounting which has been popular because of its suitability for thin films as well as bar or wire~\cite{Tri2014,Fu2017,San2018}. A recent review on the S measurement setup conclude that one has to analyze the measurement error in their setup~\cite{Che2018} for the tracking of design demerit.

In the setup of Tripathi $et~al.$~\cite{Tri2014} they claim uncertainty in the measurement of S $\sim$ 0.5~$\mu V/K$, while the sample mounting is convenient. Soni $et~al.$~\cite{Son2008} claim fast measurements with the uncertainty in S $\sim$0.22\%. Sharath $et~al.$~\cite{Sha2008} reported almost similar arrangement like Soni $et~al.$ in PPMS (physical properties measurement system) probe. In the design of Jonathan $et~al.$~\cite{Jon2010} they are able to measure from 300 K to 1273 K and the absolute error of measurement was $\sim$3-5 \%. Wood $et~al.$~\cite{Woo1985} introduced the pulse method using light pipes for small $\Delta T$. Above mentioned setups adopted the steady state or quasi steady state methods~\cite{Che2018}, however for the fast measurement speed, dynamic method have been utilized~\cite{Mun2010}. In our setup, a hybrid version of quasi steady state and dynamic method is utilized.

We have used the bridging mounting and small $\Delta T$ for negligible influence on the base temperature. We apply controlled heat pulses on one end to produce periodic $\Delta T$. For estimation of S we use the linear fit method. Fitting procedure reduces the effect of spurious or offset voltages~\cite{Che2018}. Importance of the present setup is that it can be controlled in slowest possible rate to mimic the steady state data (at cost of time) at the same time it reduces the errors from the spurious voltages. We show excellent agreement between measured and reported data of standard materials. We also show the standard uncertainty of the repeatability in S ($\sim\pm 0.056~\mu V/K$). The setup is automated with the National instruments LabView interface development for the operation and data collection.

\begin{figure*}[hbt]
\centering
\includegraphics[width=0.7\textwidth,keepaspectratio]{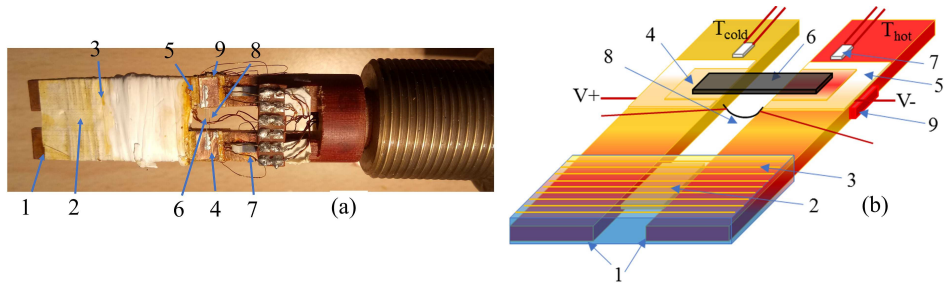}
\caption{(a) Photograph of the sample holder. (b) Temperature distribution schematic of the sample holder during the measurements. Parts marked with numbers are explained in the text.}
\label{holder}
\end{figure*}

\begin{figure*}[hbt]
\centering
\includegraphics[width=0.7\textwidth,keepaspectratio]{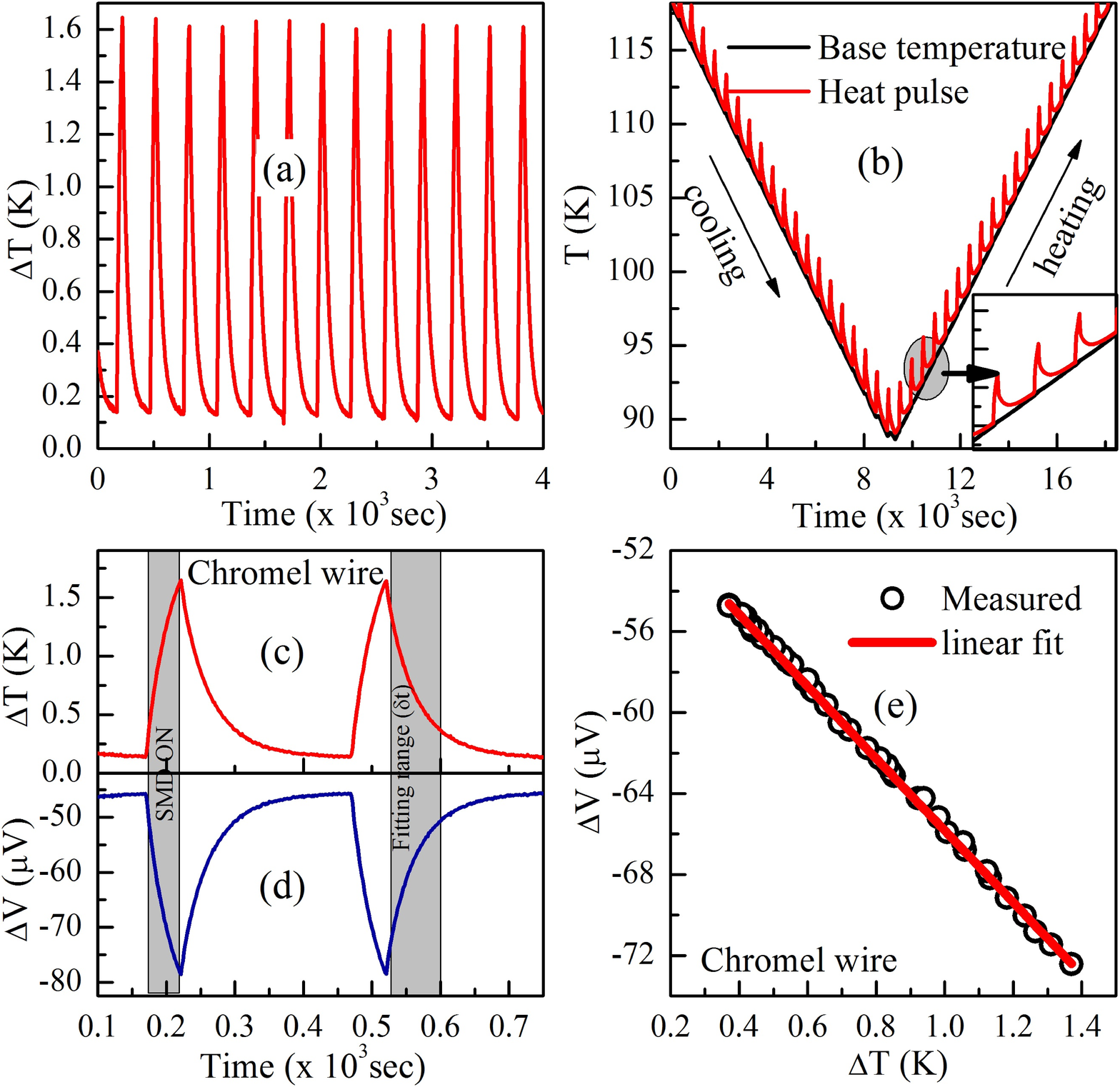}
\caption{(a) Variation of the $\Delta T$ with time. (b) Profile of base temperature during heating and cooling cycles along with $\Delta T$ profile in respective cycles. Inset shows the zoomed view of heating cycle. (c) Zoomed $\Delta T$ profile with the shaded regions when the heat pulse is ON and a typical fitting range ($\delta t$) and (d) corresponding differential voltage. (e) Linear fitting of $\Delta V$ vs $\Delta T$ for the calculation of S.}
\label{dt}
\end{figure*}

Fig.~\ref{holder} (a) displays a photograph of an optimized sample holder for producing effective temperature gradient ($\Delta T$). Now we describe the individual components in the sample holder, which are marked from 1 to 9. We have used oxygen free highly conducting (OFHC) copper bars $\left(1\right)$. Each copper bar is 50 mm long, 6 mm wide and 2 mm thick. We have inserted a piece of copper in between the bars as a heat exchanger $\left(2\right)$. At the bottom, a 50 Ohm heater $\left(3\right)$ of insulated manganin wire $\left(LakeShore\right)$ has been mounted to include both bars for the base temperature. On both sides of the bars small pieces of copper foils $\left(4\right)$ were glued through the cigarette paper $\left(5\right)$, to make electrical insulation. The sample  $\left(6\right)$ is mounted on these copper foils for measurement. Each of the copper foil piece has been mechanically anchored with the copper wire of high gauge for measuring voltages (V+/V-). Two separate Pt100 sensors $\left(7\right)$ have been employed to monitor the temperature of each bar. To measure temperature difference between the bars, on one side we have thermally anchored the differential thermocouple $\left(8\right)$ of chromel-copper. A SMD thermistor $\left(9\right)$ of 100 Ohm is glued on one of the bar through cigarette paper. This arrangement will provide identical temperature environment to both, the thermocouple and the sample under measurement. Fig.~\ref{holder} (b) shows the schematic of sample holder with the distribution of temperature on the bars during the measurement. All the wires from the sample holder have been soldered on printed circuit board (PCB) which is glued on the hylam socket, are connected through a brass vacuum feedthrough. The wires coming from the sample holder are connected to temperature controller (Cryo-con 22C) and Keithley 2182A nano-voltmeter. This whole arrangement is screwed with a brass at the bottom of the insert which is made of seamless stainless steel tube. The sample holder assembly has been covered with brass jacket for vacuum. At the time of measurement the probe will have to hang within liquid nitrogen dewar and the sample space is evacuated. 

\begin{figure*}[hbt]
\centering
\includegraphics[width=0.7\textwidth,keepaspectratio]{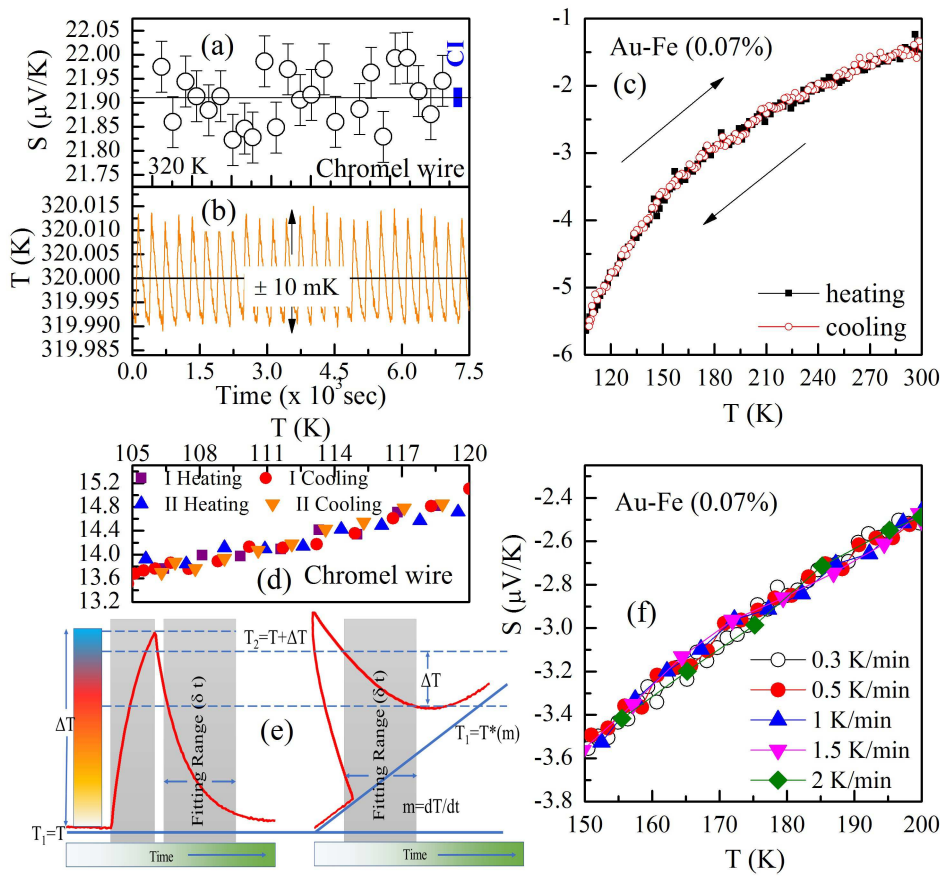}
\caption{(a) S measurement of chromel wire at fixed temperature 320 K for $>$ 10$^3$ sec, error bars shown here is the standard uncertainty, horizontal line represents the mean and blue bar shows the confidence interval (95\%). (b) Stability of base temperature at 320 K for $>$ 10$^3$ sec. (c) Measured S in heating and cooling cycle for the Au-Fe (0.07 \%). (d) Four consecutive cycles of S measurements of chromel wire in ramping mode with 0.3 K/min rate. (e) Schematics of rate of heat pulse at a constant base temperature and on the ramping base temperature (for details see text). (f) Measured S of Au-Fe (0.07 \%) with different ramp rates, from 0.3 K/min to 2 K/min.}
\label{uncer}
\end{figure*}

This setup uses only one heater for the base temperature and requires only one temperature controller. This heater is sufficient to heat the sample upto $\sim$ 400 K with the limitation of insulation of manganin heater wire. This heater can be ramped with the controlled ramp using one of the two loops of temperature controller. It makes the temperature of both the bars identical. Bar with SMD thermistor is connected with the second control loop of temperature controller and can be controlled in pulse mode. By controlling the output power and optimizing the time for the pulse to be ON, one may get the desirable $\left(\Delta T\right)$ of about 2 K to 5 K. We optimized that by applying $\sim$3 \% power $\left(75~mW\right)$ to thermistor for 50 seconds, $\Delta T$ of about 2 K can be achieved. Fig.~\ref{dt} (a) shows the profile of $\Delta T$ with time.

Our measurement program first stabilizes the set (desired) temperature and $\sim$60 sec after stabilization  a heat pulse is applied (see Fig.~\ref{dt} (c)). Heat pulse raises the temperature of the bar with SMD and a $\Delta T$ of desired value is achieved. Once desired $\Delta T$ is achieved, SMD is kept unpowered until $\Delta T$ approaches close to zero. Simultaneously $\Delta V$ of the sample is measured for each $\Delta T$ (see Fig.~\ref{dt} (d)). In nanovoltmeter, one channel measures the voltage from the sample while the other channel is dedicated to thermocouple. Time duration when $\Delta T$ decreases and approaches towards zero has been optimized for the fitting and data collection for maximum accuracy. This process provides S value for the base temperatures. The temperature increment/decrement is in the ramp mode throughout the measurement. We have used 0.3 K/min for the standard samples. Fig.~\ref{dt} (b) shows the modulation of heat pulses on base temperature in heating and cooling cycles. We have measured the Au-Fe (0.07\%) and chromel in heating-cooling cycles and excellent overlap of both the cycles is observed (see Fig.~\ref{uncer} (c) and Fig.~\ref{uncer} (d)).

\begin{figure*}[hbt]
\centering
\includegraphics[width=0.7\textwidth,keepaspectratio]{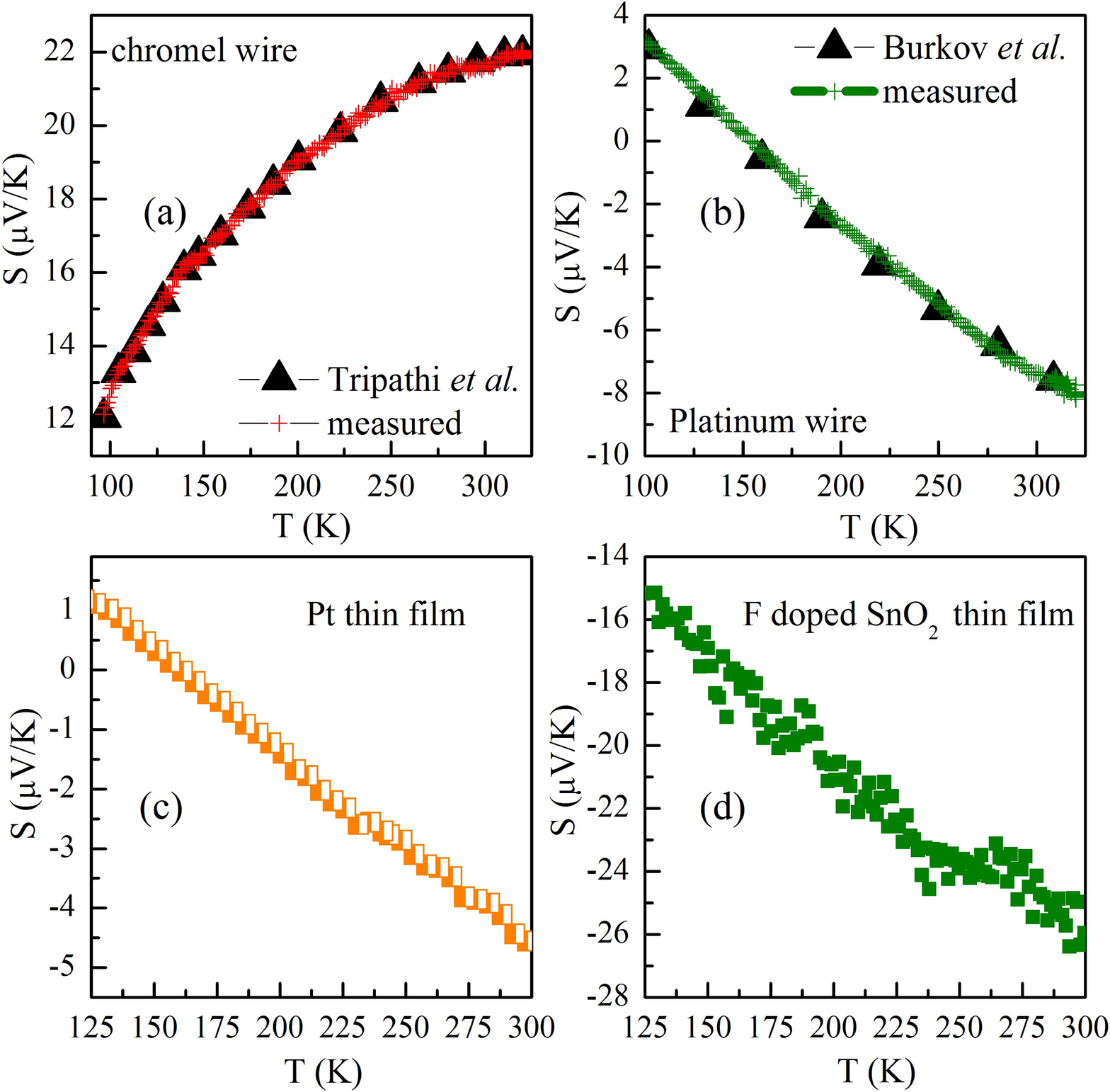}
\caption{(a) Comparison of S for chromel measured and reported data. (b) Comparison of S for Pt measured and reported data. (c) S measured for Pt thin film on quartz substrate. (d) S measured for F doped SnO$_2$ deposited on glass.}
\label{data}
\end{figure*}

Usually, the main error in S comes from the error in the measurement of $\Delta T$. To overcome this we used the thermocouple S data and measured voltage difference ($\Delta V$) to get temperature difference ($\Delta T$) through the relation~\cite{Tri2014,Fu2017} $T_{high} - T_{low} = \frac{- \Delta V}{S_{Chromel} - S_{Copper}}$. This method has been proven to reduce the error~\cite{Ash2012} in $\Delta T$. Using value of $\Delta T$ and simultaneously $\Delta V$ coming from the sample, one can calculate the value of S. For maximum accuracy, we perform linear fit (see Fig.~\ref{dt} (e)) of $\Delta V$ and $\Delta T$ during the measurement.

Fig.~\ref{uncer} (e) schematically shows the modulation of heat pulse on a stable (left) and a ramping (right) base temperature with ramp rate m (= dT/dt). One can see that the ramp rate (m) affects to the effective $\Delta T$ fitting range ($\delta t$). For different ramp rates, we can easily change $\delta t$ range by observing the linearity of ($\Delta V$ vs $\Delta T$) curve (Fig.~\ref{dt} (e)). However, the maximum rate of measurement (m) should not exceed the change in ($\Delta T$) within the fitting range ($\delta t$) ${i.e.}$ $\frac{dT}{dt} \leq \frac{\Delta T}{\delta t}$. This flexibility allows us to successfully measure the S value upto 2 K/min with good overlap (see Fig.~\ref{uncer} (f)). Faster measurement speed is suitable only if there are no phase transition, however, to observe a phase transition ramp rate must be kept slower.

In this following we will show the measured data of high purity wires of Pt, chromel and Au-Fe (0.07 \%) of high gauges. All these wires have been bridge mounted using conducting silver paint. The use of silver paint is very convenient and one may easily remove it by amyl acetate. After curing the silver paint under the heating lamp, one can start the measurement. Fig.~\ref{uncer} (a) and (b) show the data of a chromel wire sample over a large measurement time ($>$ 10$^3$ sec). From the standard statistical analysis of this data set, the standard uncertainty of the repeatability in S measurement is estimated to be  $\sim$ $\pm 0.056~\mu V/K$. We have also estimated the 95\% confidence interval (according to \textit{Guide to the expression of uncertainty in measurement}) of this data set which is represented as blue bar around the mean value in Fig.~\ref{uncer} (a). Fig.~\ref{data} (a) shows the measured S of chromel vs copper along with the reported data~\cite{Tri2014}. We observe excellent agreement with the literature. Measured data of the pure Pt wire of very high gauge ($\sim$ 30~$\mu m$) also matches well with the earlier reports~\cite{Bur2001,Tri2014} (see Fig.~\ref{data} (b)). The crossover of sign change in S for Pt also confirms the accuracy of sample T. In order to show the suitability of  our setup for the thin film S, data of a 58 nm thin Pt film on quartz substrate (deposited by magnetron sputtering) and commercially available (Techinstro, India) F doped SnO$_2$ (thickness $\sim$ 200~nm) on soda lime glass are shown in the Fig.~\ref{data} ((c) and (d)). These data confirm the suitability of setup for the S measurements of thin films.

In conclusion, we have presented an effective automated design for the S measurement for samples in form of wire, bar and thin films. Controlled heat pulse is employed to produce a temperature gradient $\Delta T$ and simultaneous $\Delta V$ measurement allows estimation of S through linear fit. The setup is able to reproduce the data in heating and cooling cycles which shows its suitability for study of thermoelectric materials which exhibits first order phase transitions. Measurement of S taken at ramp rate upto 2 K/min are shown to be overlapping with data acquired with slower rate (0.3 K/min).

We are thankful to G. S. Okram for providing high purity chromel and Au-Fe (0.07 \%) wires and M. Gupta for providing a Pt thin film. R. Rawat is gratefully acknowledged for fruitful discussions and also for providing pure Pt wire. AA acknowledges UGC, New Delhi, India for financial support (2016-17/MANF-2015-17-UTT-53853). DKS acknowledges support from DST-New Delhi, India through grant no. INI/RUS/RFBR/P-269.


\bibliography{tep_aip2}

\end{document}